\documentclass[prl, reprint, amsmath, amssymb, floatfix, aps, longbibliography, superscriptaddress]{revtex4-1}
\usepackage{graphicx}
\usepackage{dcolumn}
\usepackage{bm}
\usepackage[colorlinks=true,urlcolor=rossos,anchorcolor=black,citecolor=mygreen,filecolor=black,linkcolor=black,menucolor=blue,linktocpage=true]{hyperref} 
\usepackage{subfigure}

\usepackage{booktabs}
\usepackage{colortbl}
\usepackage{collcell}
\usepackage{enumerate}
\usepackage{float}
\usepackage{verbatim}
\usepackage{multirow}
\usepackage{xparse}
\usepackage{tabularx}
\usepackage{cancel}

\usepackage{multirow}

\usepackage[normalem]{ulem}
\usepackage{soul}
\setstcolor{red}

\definecolor{rossos}{cmyk}{0,1,1,0.55}
\definecolor{mygreen}{rgb}{0.27, 0.64, 0.48}
\definecolor{mygray}{gray}{0.95}

\begin{document}

\title{Paleodetector Bounds on Cosmic-Ray-Boosted Dark Matter}

\author{Jin-Wei Wang}
\email{jinwei.wang@uestc.edu.cn}
\affiliation{School of Physics, University of Electronic Science and Technology of China, Chengdu 611731, China}

\author{Fei-Fei Li}
\affiliation{School of Physics, University of Electronic Science and Technology of China, Chengdu 611731, China}

\begin{abstract}
We present the first dedicated analysis of cosmic-ray boosted dark matter (CRDM) in paleodetectors. 
Owing to their large kinetic energies, CRDM particles generate nuclear-recoil tracks that extend to substantially larger lengths than those produced by dominant backgrounds from neutrinos and intrinsic radioactivity. 
Combined with the ultra-large effective geological exposure of $\mathcal{O}(10^{5})~\mathrm{t\,yr}$, paleodetectors provide high sensitivity to sub-GeV DM. 
Considering both constant and vector-mediator interactions, we find that paleodetectors improve the sensitivity to the DM--proton scattering cross section by one to two orders of magnitude compared with the latest XENONnT limits.

\end{abstract}

\maketitle

\noindent \textit{\textbf{Introduction.}}---According to a broad range of astrophysical and cosmological observations, dark matter (DM) constitutes approximately 85\% of the total matter content in the universe \cite{hep-ph/0404175,Taoso:2007qk,Bertone:2016nfn,1807.06209,Arbey:2021gdg,Cirelli:2024ssz}. Despite its overwhelming abundance, the fundamental particle nature of DM remains unknown, rendering its identification one of the central challenges in contemporary particle physics.
Under the assumption that DM interacts weakly with standard model (SM) particles, direct detection experiments provide a powerful and model-independent approach to probing its properties through nuclear and/or electron recoils. Leading experiments such as LUX-ZEPLIN (LZ) \cite{LZ:2015kxe,LZ:2018qzl,LZ:2022lsv}, XENONnT \cite{XENON:2022ltv,XENON:2023cxc}, and PandaX-4T \cite{PandaX:2018wtu,PandaX-4T:2021bab} have achieved unprecedented sensitivities to weakly interacting massive particles (WIMPs) with masses in the GeV--TeV range. However, for sub-GeV and lighter DM, the sensitivity of conventional experiments rapidly deteriorates: halo DM in this mass regime carries insufficient kinetic energy to produce detectable recoils in detectors with typical thresholds at the $\mathcal{O}(\mathrm{keV})$ scale.

To overcome this limitation, several scenarios involving ``boosted'' DM populations have been proposed \cite{1405.7370,1506.04316,1708.03642,1709.06573,Bringmann:2018cvk,Wang:2021jic,Granelli:2022ysi}. A particularly robust and unavoidable mechanism was identified in Ref.~\cite{Bringmann:2018cvk}, where a small fraction of the Galactic DM halo is accelerated to relativistic energies through elastic scattering with high-energy cosmic rays (CRs). This component, commonly referred to as cosmic-ray boosted DM (CRDM), acquires sufficient kinetic energy to generate observable signals in terrestrial detectors, thereby enabling sensitivity to DM--proton interactions well below the GeV mass scale.

In this work, we demonstrate that ancient minerals---so-called paleodetectors---provide high sensitivity to CRDM. The detection principle of paleodetectors is straightforward: a DM particle scattering off a nucleus in a crystalline mineral produces a nuclear recoil, which subsequently traverses the lattice and deposits energy through ionization losses. In insulating materials, this process leaves permanent damage tracks that can be resolved with modern microscopy techniques at nanometer-scale spatial resolution. The resulting track length and morphology distributions thus provide a direct record of the underlying DM–nucleus interaction and recoil kinematics. 

Paleodetectors constitute a fundamentally different detection paradigm: rather than operating a detector in real time, one searches for the permanent damage tracks imprinted by nuclear recoils in natural minerals over geological timescales, up to $\mathcal{O}(1)\,\mathrm{Gyr}$. 
A key advantage of paleodetectors lies in their enormous effective exposure. For example, 100 g of material that has been recording nuclear damage tracks for 1 Gyr yields an effective $\mathcal{O}(10^5)~\mathrm{t\,yr}$ exposure, which is far beyond what is achievable in conventional experiments with $\mathcal{O}(1)~\mathrm{t\,yr}$ exposure. This renders paleodetectors particularly well suited for probing extremely feeble interactions, such as DM \cite{Snowden-Ifft:1995zgn,Baum:2018tfw,Drukier:2018pdy,Edwards:2018hcf,Baum:2021jak,Baum:2021chx,Bramante:2021dyx,Baum:2023cct,Fung:2025cub} and neutrinos \cite{Baum:2019fqm,Jordan:2020gxx,Tapia-Arellano:2021cml,Baum:2022wfc,Baum:2023cct}.

Here, we present the first dedicated analysis of CRDM in paleodetectors. We demonstrate that their sensitivities to the DM--proton scattering cross section surpass those of existing direct detection experiments by one to two orders of magnitude, depending on the underlying DM model. Our results establish paleodetectors as a powerful and complementary probe of sub-GeV DM and highlight their unique potential to explore regions of parameter space that are otherwise inaccessible to current and near-future experiments.\\
%Besides, to reach the neutrino floor, the total exporsure shoulb be $\mathcal{O}(1000)$ t$\cdot$y \cite{XLZD:2024nsu}.

\noindent \textit{\textbf{CRDM flux.}}---
For the interaction between DM and protons, we consider two benchmark scenarios. 
The first is a phenomenological model with a constant DM--proton scattering cross section, denoted by $\sigma_{\chi p}$. 
The second is a simplified model with a massive vector mediator, described by the Lagrangian \cite{Guo:2020oum,Su:2023zgr,Wang:2025ztb}
\begin{equation}
\begin{split}
\mathcal{L} &\supset \bar{\chi}\left(i\gamma^\mu\partial_\mu - m_\chi\right)\chi + g_\chi \bar{\chi}\gamma^\mu \chi V_\mu \\
&\quad + \sum_q g_q \bar{q}\gamma^\mu q V_\mu + \frac{1}{2} m_V^2 V_\mu V^\mu ,
\end{split}
\end{equation}
where $m_\chi$ and $m_V$ denote the masses of the DM particle and the vector mediator, respectively, and $g_\chi$ and $g_q$ are the corresponding couplings to DM and quarks. 
Throughout this work we assume flavor-universal couplings between the mediator and quarks. 
Extensions to other interaction structures, such as scalar or axial-vector mediators, are straightforward and will not qualitatively alter our conclusions \cite{DeMarchi:2024riu,Su:2022wpj,DeMarchi:2025uoo,DeMarchi:2025xag}.

The local interstellar (LIS) population of CRs is well measured and is commonly characterized by their differential intensity $dI/dR$ as a function of rigidity $R$. 
In this work, we adopt the LIS spectra from Ref.~\cite{Boschini:2020jty}, where Galactic CR propagation and solar modulation are treated using \texttt{GALPROP} and \texttt{HELMOD}, respectively. 
These results are consistent with Voyager~1 observations \cite{Cummings:2016pdr} as well as AMS-02 measurements \cite{AMS:2020cai}. 

Cosmic-ray scattering on halo DM unavoidably produces a boosted DM component. 
The resulting differential CRDM flux can be written as
\begin{equation}
\frac{d\Phi_\chi}{dT_\chi}
= D_{\rm eff}\,\frac{\rho_\chi}{m_\chi}
\sum_i \int^\infty_{T_i^\text{min}(T_\chi)} dT_i \,
\frac{d\sigma_{\chi i}}{dT_\chi}\,
\frac{d\Phi_i}{dT_i}
\label{eq:CRDMflux}
\end{equation}
with
\begin{equation}
    T_i^{\text{min}}(T_\chi) = \left( \frac{T_\chi}{2} - m_i \right) 
\left[ 1 \pm \sqrt{1 + \frac{2T_\chi}{m_\chi} 
\frac{(m_i + m_\chi)^2}{(2m_i - T_\chi)^2}} \right],
\label{eq:Tklimits}
\end{equation}
where the $+$ ($-$) sign applies for $T_\chi > 2m_i$ ($T_\chi < 2m_i$), $\rho_\chi \simeq 0.3~{\rm GeV\,cm^{-3}}$ is the local DM density, $T_\chi$ ($T_i$) denotes the kinetic energy of DM (CR species $i$), and $d\sigma_{\chi i}/dT_\chi$ is the differential scattering cross section between DM and CR species $i$. 
The parameter $D_{\rm eff}$ represents the effective distance over which CR--DM scattering contributes to the local CRDM flux; throughout this work we take $D_{\rm eff}\simeq 9~{\rm kpc}$\footnote{The astrophysical uncertainties associated with CRDM production, including CR propagation, the spatial distribution of CR sources, and the Galactic DM profile, are effectively encoded in the parameter $D_{\rm eff}$~\cite{Xia:2021vbz}. Depending on $m_\chi$ and $T_\chi$, $D_{\rm eff}$ typically varies between $\sim6~{\rm kpc}$ and $\sim10~{\rm kpc}$, while different DM halo profiles modify it at the level of approximately $10\%$~\cite{Xia:2021vbz}. The NFW profile adopted in this work yields the most conservative CRDM flux. Since the CRDM flux scales linearly with $D_{\rm eff}$, the projected sensitivity approximately scales as $\sigma_{\chi p}\propto1/\sqrt{D_{\rm eff}}$. For the light and high-energy CRDM considered in this work, we adopt $D_{\rm eff}=9~{\rm kpc}$ as a benchmark value following Ref. \cite{Xia:2021vbz}.}. 
We include the nine dominant CR species that account for more than $90\%$ of the total CRDM flux, namely $\{\mathrm{H},\mathrm{He},\mathrm{C},\mathrm{N},\mathrm{O},\mathrm{Ne},\mathrm{Mg},\mathrm{Si},\mathrm{Fe}\}$ \cite{Xia:2021vbz}. 

The differential scattering cross section $d\sigma_{\chi i}/dT_\chi$ is model dependent. 
For the constant cross section scenario, we have
\begin{equation}
\frac{d\sigma_{\chi i}}{dT_\chi} = A_i^2\,\frac{G_i^2(Q^2)}{T_\chi^{\rm max}(T_i)}\,\frac{\mu_{\chi i}^2}{\mu_{\chi p}^2}\,\sigma_{\chi p},
\label{eq:constDMA}
\end{equation}
with
\begin{equation}
T_{\chi}^{\rm max}(T_i) = \frac{T_i^2 + 2m_i T_i}{T_i + (m_i + m_\chi)^2 / (2m_\chi)},
\end{equation}
where $A_i$ is the mass number of CR species $i$, $\mu_{\chi i}$ and $\mu_{\chi p}$ are the reduced masses of the DM--nucleus and DM--proton systems, respectively, and $G_i(Q^2)$ denotes the nuclear form factor with positive four-momentum transfer squared $Q^2$. 
For hydrogen and helium, we adopt the dipole form factor \cite{Angeli:2004kvy,Bringmann:2018cvk}, while for heavier nuclei we use the Helm form factor \cite{PhysRev.104.1466,Lewin:1995rx,Xia:2021vbz,Wang:2025ztb}.

For the vector-mediator scenario, both elastic and deep-inelastic scattering (DIS) processes contribute. 
The elastic DM--nucleus $i$ scattering cross section is given by \cite{Guo:2020oum,Su:2023zgr,Wang:2025ztb}
\begin{equation}
\frac{d\sigma^{\rm el}_{\chi i}}{dT_\chi}
= 2m_\chi \frac{A^2_i g_p^2 g_\chi^2}{4\pi \beta^2}
\frac{G_i^2(Q^2)}{(Q^2+m_V^2)^2}
\left[ 1 - \frac{\beta^2 Q^2}{Q^2_{\rm max}} + \frac{Q^4}{8 m_\chi^2 E_i^2} \right],
\label{eq:elDMA}
\end{equation}
where $g_p = 3 g_q$ is the effective DM--proton coupling, 
%$Q^2 \equiv -q^2$ is the positive four-momentum transfer squared, 
$\beta=|{\bf p}_i|/E_i$ is the CR's velocity. 
The maximal momentum transfer is
\begin{equation}
Q^2_{\rm max} = \frac{4(E_i^2-m_i^2)m_\chi^2}{m_\chi^2 + m_i^2 + 2 m_\chi E_i}.
\end{equation}

At sufficiently high energies, inelastic scattering becomes important. 
We treat the DIS process using the standard Bjorken variables \cite{Buckley:2014ana,AbdulKhalek:2022fyi}, following Refs.~\cite{ParticleDataGroup:2024cfk,Berger:2018urf,Su:2023zgr,Wang:2025ztb,DeMarchi:2025uoo}. 
The differential cross section is
\begin{equation}
\frac{d^2\sigma^{\rm DIS}_{\chi p}}{dT_\chi d\mu_s}
= \frac{m_\chi\sqrt{(T_p^2+2m_pT_p)(T_\chi^2+2m_\chi T_\chi)}}{m_p^2 E_\chi^2 y}
\frac{d^2\sigma_{\rm DIS}}{dxdy},
\label{eq:ddxsec_dis}
\end{equation}
where $\mu_s$ is the cosine of the scattering angle, and $x$ and $y$ are the Bjorken variables. 
The partonic cross section reads
\begin{equation}
\begin{split}
\frac{d^2\sigma_{\rm DIS}}{dxdy} &=
\frac{g_\chi^2 g_q^2}{4\pi(Q^2+m_V^2)^2}
\frac{2 E_\chi^3 m_p}{E_\chi^2-m_\chi^2} \\
&\times\left(1-y+\frac{y^2}{2} - \frac{xy m_p}{2 E_\chi} -\frac{m_\chi^2}{2 E_\chi m_p} \frac{y}{x}\right)F_2(x, Q^2),
\end{split}
\end{equation}
with the structure function
\begin{equation}
F_2(x,Q^2) = \sum_{f=u,d,c,s,b} \left[f(x,Q^2) + \bar f(x,Q^2)\right] x,
\end{equation}
where $f$ ($\bar f$) are the parton distribution functions (PDFs) of quarks (antiquarks). 
We use the MSTW 2008 NNLO PDFs \cite{Martin:2009iq}. 
For scattering on CR nuclei, we approximate
\begin{equation}
\frac{d^2\sigma^{\rm DIS}_{\chi i}}{dT_\chi d\mu_s}
= R_i(x,Q^2)\left[Z_i\frac{d^2\sigma^{\rm DIS}_{\chi p}}{dT_\chi d\mu_s}+(A_i-Z_i)\frac{d^2\sigma^{\rm DIS}_{\chi n}}{dT_\chi d\mu_s}\right],
\end{equation}
where $R_i(x,Q^2)$ denotes the nuclear modification factor, which is $\mathcal{O}(1)$ \cite{Kovarik:2015cma,Eskola:2016oht}. 
For simplicity, we set $R_i(x,Q^2)\simeq1$ in our numerical analysis. 
Integrating over the scattering angle, we obtain
\begin{equation}
\frac{d\sigma^{\rm DIS}_{\chi i}}{dT_\chi}
=\int d\mu_s \,\frac{d^2\sigma^{\rm DIS}_{\chi i}}{dT_\chi d\mu_s}.
\label{eq:disDMA}
\end{equation}

Combining Eqs.~(\ref{eq:CRDMflux})--(\ref{eq:disDMA}), we compute the CRDM flux for the different interaction scenarios. 
We focus on the light DM regime, $m_\chi \lesssim 0.1~{\rm GeV}$, which is particularly well motivated for boosted DM searches. 
For the mediator mass we consider two benchmark values, $m_V=1~{\rm GeV}$ and $m_V=10~{\rm MeV}$; results for other values of $m_V$ can be inferred from these benchmarks.
\begin{figure}
\centering
\includegraphics[width=0.45\textwidth]{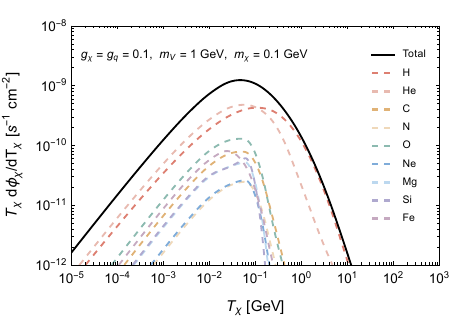}\\
\vspace{0.3cm}
\includegraphics[width=0.45\textwidth]{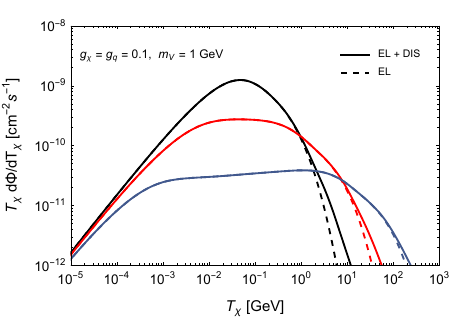}
\caption{Differential flux of CRDM for the vector-mediator model. 
(Top) Total CRDM spectrum (solid black) and the individual contributions from different cosmic-ray species (dashed), assuming $g_\chi=g_q=0.1$, $m_V=1~{\rm GeV}$, and $m_\chi=0.1~{\rm GeV}$. 
(Bottom) Impact of inelastic scattering on the CRDM flux: the solid curves include both elastic and deep-inelastic scattering (EL+DIS), while the dashed curves show the elastic-only contribution (EL). 
Different colours correspond to different DM mass $m_\chi$, namely $m_\chi=100~{\rm MeV}$ (black), $10~{\rm MeV}$ (red), and $1~{\rm MeV}$ (blue).}
\label{fig:DM_Tx_Spectrum}
\end{figure}

In Fig.~\ref{fig:DM_Tx_Spectrum}, we show the resulting CRDM spectra for the vector-mediator model. 
The upper panel displays the total CRDM flux (solid black), together with the individual contributions from different CR species (dashed). 
The lower panel illustrates the impact of inelastic scattering: the solid and dashed curves correspond to the full calculation (including both elastic and DIS processes) and to the elastic-only case, respectively. 
We consider three representative DM masses, $m_\chi=100~{\rm MeV}$ (black), $10~{\rm MeV}$ (red), and $1~{\rm MeV}$ (blue), and fix $g_\chi=g_q=0.1$ with $m_V=1~{\rm GeV}$. 
As is evident from the figure, the DIS contribution becomes increasingly important at high energies, demonstrating that inelastic processes must be included for an accurate description of the CRDM flux. For comparison, the standard halo DM flux is approximately
$
\Phi_\chi^{\rm halo}\sim
\frac{\rho_\chi \bar v_\chi}{m_\chi}
\simeq
9\times10^8
\left(
\frac{10~{\rm MeV}}{m_\chi}
\right)
{\rm cm^{-2}\,s^{-1}},
$
assuming $\rho_\chi=0.3~{\rm GeV\,cm^{-3}}$ and $\bar v_\chi\simeq10^{-3}c$. Integrating the differential CRDM spectrum shown in the top panel of Fig.~1 gives
$\Phi_\chi^{\rm CRDM}\simeq4.82\times10^{-9}~{\rm cm}^{-2}{\rm s}^{-1}$,
showing that the boosted component constitutes only a tiny fraction of
the total Galactic DM population.\\

\noindent\textit{\textbf{CRDM track length spectrum.}}—
For both conventional direct detection experiments and paleodetectors, a key physical quantity is the CRDM-induced nuclear recoil rate on a target nucleus $j$ with mass $m_j$. 
In direct detection experiments, $j$ typically corresponds to xenon (Xe), whereas in paleodetectors it denotes the various constituent nuclei of the mineral. 
Since both techniques are sensitive only to relatively low nuclear recoil energies, $T_j \lesssim \mathcal{O}(0.1)\,\mathrm{GeV}$, we restrict our analysis to elastic CRDM--nucleus scattering.

Using Eq.~(\ref{eq:CRDMflux}), the differential recoil rate is
\begin{equation}
\frac{d\Gamma_j}{dT_j}
= \int_{T_\chi^{\rm min}(T_j)}^{\infty} dT_\chi \,
\frac{d\sigma_{\chi j}}{dT_j}\,
\frac{d\Phi_\chi}{dT_\chi},
\label{eq:gammarate}
\end{equation}
where $d\sigma_{\chi j}/dT_j$ and the minimal CRDM energy $T_\chi^{\rm min}(T_j)$ follow from Eqs. \eqref{eq:elDMA}/\eqref{eq:constDMA} and \eqref{eq:Tklimits}, with the replacements $\chi \to j$ and $i \to \chi$.

The CRDM signal in paleodetectors is encoded in the spectrum of damage track lengths. 
The differential rate of tracks with reconstructed length $x_T$ for an exposure $\epsilon$ is given by \cite{Baum:2018tfw,Drukier:2018pdy,Edwards:2018hcf,Baum:2021jak}
\begin{equation}
\frac{dR}{dx_T}
=\epsilon \sum_j \frac{\xi_j}{m_j}\,
\frac{d\Gamma_j}{dT_j}\,
\frac{dT_j}{dx_T},
\label{eq:dRdx}
\end{equation}
where $\xi_j$ denotes the mass fraction of element $j$ in the composite target, and $dT_j/dx_T$ is the stopping power, evaluated using \texttt{SRIM} \cite{10.1007/978-3-642-68779-2_5,Ziegler:2010bzy}. 
Ions with charge $Z\leq2$ (in particular H and He) do not produce persistent tracks in most materials and are therefore neglected \cite{Drukier:2018pdy,Baum:2021jak}. Note that at high recoil energies, electronic stopping (ES) may modify the relation
between recoil energy and observable track length. Recent TRIM
simulations indicate a suppression of the growth of characteristic track
lengths once ES becomes dominant~\cite{Fung:2025cub}, although the detailed
microscopic relation between stopping mechanisms and persistent track
formation remains under investigation. To estimate its impact, we adopt a
conservative phenomenological truncation prescription at the crossover
between nuclear and ES powers. A microscopic treatment of track formation in this regime is beyond the
scope of the present work. We use this prescription to assess the possible
impact of ES on our sensitivity projections; the
quantitative results are presented below in Fig.~\ref{fig:DM_constraints}.

Two complementary readout strategies have been proposed for paleodetectors \cite{Baum:2018tfw,Drukier:2018pdy,Edwards:2018hcf,Baum:2021jak}. 
The high-resolution scenario probes small samples (e.g., $\sim10$\,mg) with nanometer-scale track length resolution, while the high-exposure scenario analyzes much larger samples (e.g., $\sim100$\,g) with coarser resolution (e.g., $\sim15$\,nm). 
In this work, we adopt the latter, which maximally exploits the large CRDM-induced track lengths and the ultra-large effective exposure.
%\footnote{\textcolor{blue}{Recent experimental progress toward large-volume paleodetector technologies has been reported in Ref.~\cite{Araujo:2025rhr}, where three-dimensional nuclear recoil tracks were imaged deep inside cubic-centimeter-sized LiF crystals using recoil-induced color centers and light-sheet fluorescence microscopy. The authors pointed out that these results constitute the first step toward future $10$--$1000~{\rm g}$ scale detectors with single-event sensitivity for applications such as dark matter and neutrino detection.}}

We emphasize that the high-exposure scenario adopted here represents a
prospective sensitivity benchmark rather than a presently demonstrated
experimental capability. In particular, simultaneously achieving a large
target mass and nanometer-scale spatial resolution, while maintaining
long-term track preservation and reliable background discrimination, remains
experimentally challenging. No mature technique has yet demonstrated all of
these capabilities at the scale assumed in our projections. Nevertheless,
ongoing proof-of-principle developments in nanoscale imaging and recoil-track
readout provide encouraging progress toward future large-volume paleo
detectors \cite{Holler:2014,Rodriguez:2014,Schaff:2015,Calabrese-Day:2026soq}. We further note that the adopted exposure is a benchmark parameter.
Since CRDM production and subsequent detection each introduce one power
of $\sigma_{\chi p}$, the signal yield scales approximately as
$\epsilon\sigma_{\chi p}^{2}$. Consequently, when the modeled background contribution in the long-track
region is negligible, the projected cross-section sensitivity
scales as $\sigma_{\chi p}^{\rm lim}\propto\epsilon^{-1/2}$.
Other exposure assumptions can therefore be readily accommodated within
the same framework.

The experimental observable is the binned track length spectrum. 
The expected number of events in the $i$-th bin, $x_T\in[x_a,x_b]$, is
\begin{equation}
N_i(x_a,x_b)
= \int_0^\infty dx_T\,
W(x_T;x_a,x_b)\,
\frac{dR}{dx_T},
\label{eq:Nevents}
\end{equation}
with the window function
\begin{equation}
W(x_T;x_a,x_b)
=\frac{1}{2}\!\left[
\mathrm{erf}\!\left(\frac{x_T-x_a}{\sqrt{2}\sigma_{x_T}}\right)
-\mathrm{erf}\!\left(\frac{x_T-x_b}{\sqrt{2}\sigma_{x_T}}\right)
\right],
\end{equation}
which accounts for the finite readout resolution $\sigma_{x_T}$. 
We consider two representative minerals, Gypsum [$\mathrm{Ca(SO_4)\!\cdot\!2(H_2O)}$] and Olivine [$\mathrm{Mg}_{1.6}\mathrm{Fe}^{2+}_{0.4}(\mathrm{SiO}_4)$], and adopt a benchmark exposure of $100~\mathrm{g\,Gyr}$ \cite{Baum:2018tfw,Drukier:2018pdy,Edwards:2018hcf,Baum:2021jak,Baum:2021chx,Bramante:2021dyx,Baum:2023cct,Fung:2025cub}. 

Only nuclear recoils constitute relevant backgrounds for paleodetectors, since electron recoils do not leave stable tracks. 
The dominant background sources are neutrinos (solar, atmospheric, and supernova) \cite{OHare:2020lva,Baum:2019fqm}, radiogenic neutrons from spontaneous fission and $(\alpha,n)$ reactions associated with the $^{238}{\rm U}$ decay chain, and recoils induced by $^{234}$Th \cite{Collar:1995aw,Snowden-Ifft:1996dug}. 
For the fiducial $^{238}\mathrm{U}$ concentrations, we adopt the same benchmark values as in Refs.~\cite{Baum:2018tfw,Drukier:2018pdy,Edwards:2018hcf,Baum:2021jak}, namely $10^{-11}$ for Gypsum and $10^{-10}$ for Olivine. Additional backgrounds, including intrinsic crystal defects, direct
spontaneous fission events, and completed $^{238}{\rm U}$ decay chains,
generally exhibit characteristic morphologies or long-/multiple-track
signatures distinct from isolated DM-induced nuclear recoils. Their practical
rejection efficiency, however, depends on the specific readout technique. More detailed discussions of the relevant backgrounds can be found in Refs.~\cite{Drukier:2018pdy,ejm-33-249-2021,Baum:2019fqm}. Signal and background spectra are computed using \texttt{paleoSpec}~\cite{Baum:2018tfw,Drukier:2018pdy,Baum:2019fqm,Baum:2021chx}.
\begin{figure}
\centering
\includegraphics[width=0.46\textwidth]{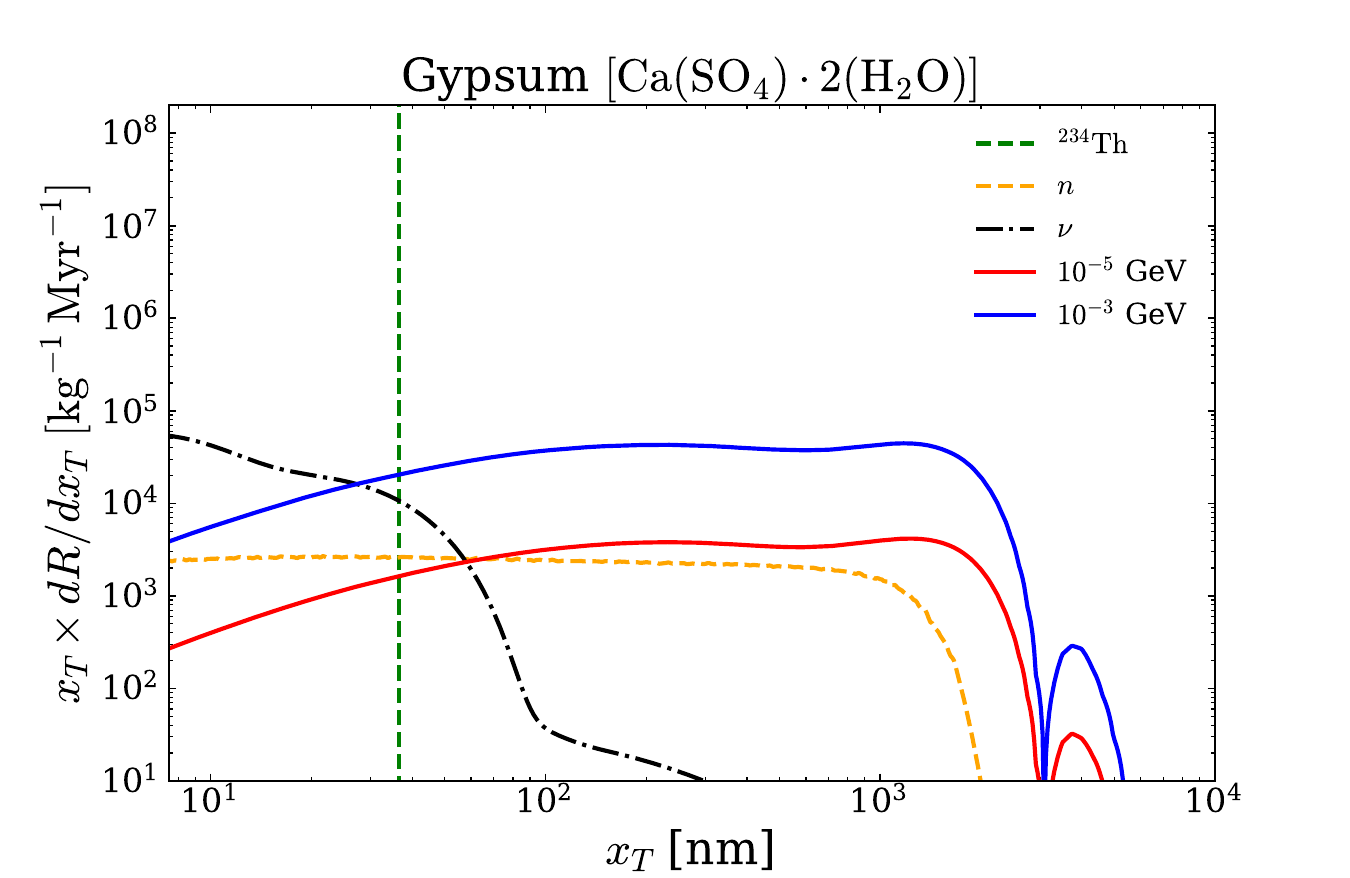}\\
\vspace{0.3cm}
\includegraphics[width=0.46\textwidth]{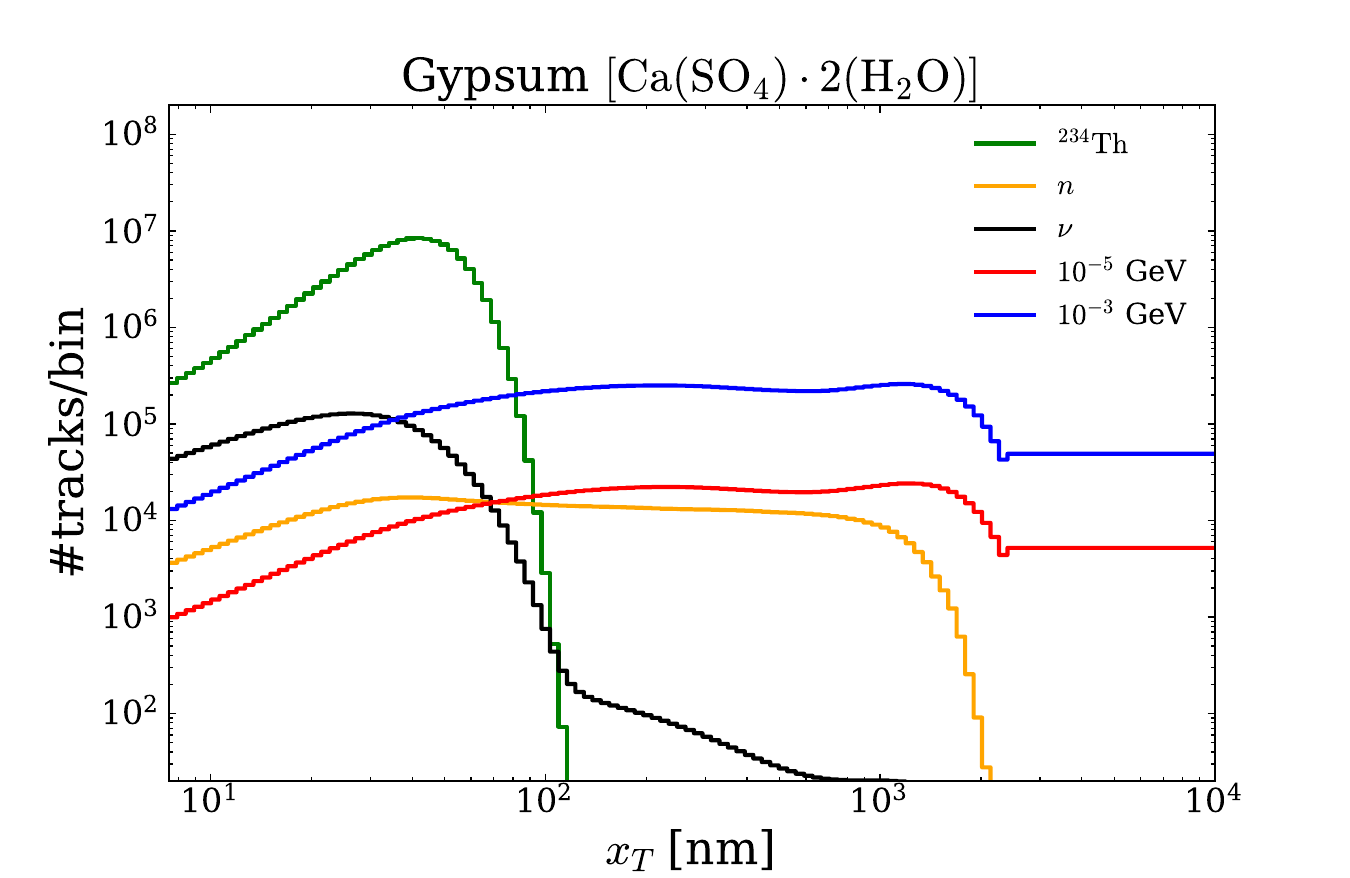}
\caption{Continuous (top) and binned (bottom) track-length distributions in
Gypsum [$\mathrm{Ca(SO_4)\cdot 2(H_2O)}$] for CRDM in the vector-mediator model.
The red and blue curves show CRDM signals for $m_\chi=10^{-5}~\mathrm{GeV}$ and $10^{-3}~\mathrm{GeV}$, respectively, while the black, orange, and green curves denote the dominant nuclear recoil backgrounds from neutrinos, radiogenic neutrons, and $^{238}\mathrm{U}\!\rightarrow\!^{234}\mathrm{Th}+\alpha$ decays. 
The input parameters are $g_\chi=g_q=0.7$ and $m_V=1~\mathrm{GeV}$.}
\label{fig:track_spectrum}
\end{figure}

Fig.~\ref{fig:track_spectrum} shows the continuous and binned track-length distributions in
Gypsum for the vector-mediator model with $g_\chi=g_q=0.7$ and
$m_V=1~{\rm GeV}$. The upper panel shows the continuous track-length
spectrum, $x_T dR/dx_T$, obtained from Eq.~\eqref{eq:dRdx}, while the lower panel
shows the corresponding binned event distribution constructed according
to Eq.~\eqref{eq:Nevents} and used in the statistical analysis. The red and blue curves
correspond to $m_\chi=10^{-5}~{\rm GeV}$ and
$m_\chi=10^{-3}~{\rm GeV}$, respectively.
A salient feature is the pronounced excess of CRDM-induced events at large track lengths, where all relevant backgrounds rapidly fall off. 
This long track regime, which is inaccessible to standard halo DM, provides a potentially powerful spectral handle for background
discrimination.

An additional important feature is the construction of the highest track-length bin. We choose the upper edge of this bin at $x_T=10~\mu{\rm m}$ and determine its lower boundary such that the expected number of neutron-induced background events is below $0.5$, thereby defining a
neutron-suppressed long-track region.
We note that the nonvanishing event number in this bin does not imply the existence of recoil tracks extending all the way to $10~\mu{\rm m}$. Rather, the final bin covers a broad interval in track length, approximately from $2000~{\rm nm}$ to $10^4~{\rm nm}$, while the underlying continuous track-length spectrum already effectively terminates around $\sim5000~{\rm nm}$. We have also verified that varying the upper edge of the final bin between $5000~{\rm nm}$ and $10^4~{\rm nm}$ has a negligible impact on the final projected sensitivities.

To summarize, our sensitivity forecast adopts a prospective high-exposure
benchmark consisting of a $100~{\rm g}$ mineral sample with an age of
$1~{\rm Gyr}$ and a track-length resolution of approximately
$15~{\rm nm}$. Persistent recoil tracks from nuclei with $Z>2$ are assumed
to be reconstructed with unit efficiency within the analyzed track-length
range, including the approximately $2$--$10~\mu{\rm m}$ interval
corresponding to the highest track-length bin in our analysis. Track annealing over the sample
age is neglected. Neutrino, radiogenic-neutron, and $^{234}{\rm Th}$ recoil
backgrounds are included explicitly in the spectral analysis, while additional
backgrounds with characteristic long- or multiple-track signatures, such as
direct spontaneous-fission events, are assumed to be distinguishable with an
appropriate readout technique. These assumptions define the benchmark
adopted in the sensitivity analysis below.\\

\noindent\textit{\textbf{Paleodetector constraints.}}—
For the vector-mediator scenario, to facilitate comparison with the literature,  we define the DM--proton scattering cross section as
\begin{equation}
\sigma_{\chi p} = \frac{g_{\chi}^2 g_p^2}{\pi m_V^4}\,\mu_{\chi p}^2,
\end{equation}
where $\mu_{\chi p} = m_\chi m_p/(m_\chi + m_p)$ is the reduced mass. 
All experimental sensitivities are presented in the $m_\chi$–$\sigma_{\chi p}$ plane.

To derive the projected paleodetector sensitivities, we analyze the
binned track-length spectra using \texttt{paleoSens}~\cite{Baum:2018tfw,Drukier:2018pdy,Baum:2019fqm,Baum:2021chx} and
follow the profile-likelihood procedure of Ref.~\cite{Baum:2021jak}. For each
track-length bin $i$, the expected event number is the sum of the CRDM
signal and background contributions. Assuming Poisson statistics for the
binned event counts, we construct a likelihood in which the background
normalizations are treated as nuisance parameters with Gaussian
constraints. We adopt a background-only Asimov data set and construct the
profile-likelihood-ratio test statistic $q(\sigma_{\chi p})$. For each fixed
$m_\chi$, the projected 90\% C.L.\ upper limit on $\sigma_{\chi p}$ is
determined by $q(\sigma_{\chi p})=2.71$~\cite{Baum:2021jak}.

For comparison, we also derive constraints from conventional direct detection using the latest XENONnT data with an exposure of $\epsilon=3.1~{\rm t\,yr}$ \cite{XENON:2025vwd}. Following Ref.~\cite{XENON:2025vwd}, we adopt the nominal background model and combine the SR0, SR1a, and SR1b data sets in a total-count analysis. As reported in Table~I of Ref.~\cite{XENON:2025vwd}, the total observed number of events is 957, while the corresponding expected background is 962 events. Assuming Poisson statistics for the total event counts \cite{ParticleDataGroup:2020ssz}, we derive the 90\% C.L.\ upper limit on the number of CRDM-induced signal events to be 35.87.

The expected number of CRDM events in XENONnT is
\begin{equation}
N_{\rm CRDM}
=\frac{\epsilon}{m_{\rm Xe}}
\int_{T_{\rm exp}^{\rm min}}^{T_{\rm exp}^{\rm max}} dT_{\rm Xe}\,
\frac{d\Gamma_{\rm Xe}}{dT_{\rm Xe}},
\end{equation}
where  $[T_{\rm exp}^{\rm min},T_{\rm exp}^{\rm max}]=[3.8,64.1]~{\rm keV}$ is energy range of sensitivity of XENONnT. 
The corresponding exclusion is obtained by imposing $N_{\rm CRDM}<N_{\rm lim}$.

\begin{figure}
\centering
\includegraphics[width=0.46
\textwidth]{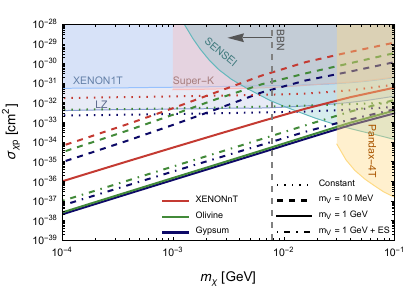}
\caption{Constraints on the DM--proton scattering cross section as a function of the DM mass. 
The purple and green curves show the projected sensitivities of paleodetectors using Gypsum and Olivine, respectively, while the red curves indicate the XENONnT constraints on CRDM. 
The dotted curves correspond to the constant cross section scenario, whereas the dashed and solid curves denote the vector-mediator model with $m_V=10~\mathrm{MeV}$ and $m_V=1~\mathrm{GeV}$, respectively. The dot-dashed curves show the corresponding $m_V=1~{\rm GeV}$
sensitivities when ES is included. Existing CRDM constraints from XENON1T \cite{Bringmann:2018cvk}, Super-K \cite{Super-Kamiokande:2022ncz}, and LZ \cite{LZ:2025iaw}, together with complementary
constraints from PandaX-4T~\cite{PandaX:2023xgl}, SENSEI~\cite{SENSEI:2023zdf}, and BBN~\cite{Giovanetti:2021izc}, are also shown for comparison. The left-pointing arrow indicates the region excluded by BBN.}
\label{fig:DM_constraints}
\end{figure}

Fig. \ref{fig:DM_constraints} summarizes the resulting limits on $\sigma_{\chi p}$. 
The dotted curves correspond to the constant cross section scenario, while the dashed and solid curves denote the vector-mediator model with $m_V = 10~{\rm MeV}$ and $1~{\rm GeV}$, respectively. 
The dot-dashed curves show the corresponding $m_V=1~{\rm GeV}$
sensitivities when ES is included.
The purple and green lines show the projected paleodetector sensitivities for Gypsum and Olivine, while the red curves indicate the XENONnT constraints. 
Existing CRDM constraints from XENON1T \cite{Bringmann:2018cvk}, Super-K \cite{Super-Kamiokande:2022ncz}, and LZ \cite{LZ:2025iaw}, together with complementary constraints from PandaX-4T~\cite{PandaX:2023xgl}, SENSEI \cite{SENSEI:2023zdf}, and BBN \cite{Giovanetti:2021izc}, are also shown for comparison. For the BBN constraint, the region toward lower $m_\chi$, indicated by the left-pointing arrow, is excluded.

As shown in Fig. \ref{fig:DM_constraints}, paleodetectors remain highly competitive in the sub-GeV mass range and can improve existing direct detection sensitivities by up to one to two orders of magnitude, depending on the interaction model. 
Even compared with the latest LZ results~\cite{LZ:2025iaw}, the projected paleodetector sensitivities remain comparable to current limits and are even stronger in part of the parameter space. 
The enhanced sensitivity mainly originates from the ultra-long exposure
times of paleodetectors and from the long-track region in which the modeled
backgrounds rapidly decrease. 

ES moderately weakens the projected sensitivities. Its
impact is largest for the $m_V=1~{\rm GeV}$ benchmark, where the limits are
weakened by approximately a factor of two, while the corresponding changes
for the constant cross section scenario and the $m_V=10~{\rm MeV}$ case
are smaller. Thus, ES does not qualitatively alter our
conclusions.\\

\noindent\textit{\textbf{Conclusion.}}—
We have presented the first dedicated study of CRDM in paleodetectors. 
We considered two benchmark DM--proton interaction scenarios, namely a constant cross section model and a vector-mediator model, and incorporated the dominant cosmic-ray contributions relevant for CRDM production.

Owing to the high kinetic energies of CRDM particles, the induced nuclear-recoil tracks extend to substantially larger lengths than those produced by dominant backgrounds from neutrinos and intrinsic radioactivity. 
Together with their ultra-large effective geological exposure of $\mathcal{O}(10^{5})~\mathrm{t\,yr}$, this enables sensitivities to the DM--proton scattering cross section that improve by up to two orders of magnitude compared with current direct detection limits, including those from XENONnT.

Our results highlight the potential of paleodetectors as a powerful and
complementary future probe of boosted DM and motivate further theoretical
and experimental studies of geological materials for DM detection.\\

%%%%%%%%%%%%%%%%%%%%%%%%%%%%

\noindent \textit{\textbf{Acknowledgments.}}---The authors wish to thank Alessandro Granelli for the thoughtful discussions at the initial stage of this project. The work of  J.-W.W. was supported by the National Natural Science Foundation of China (NSFC) under Grants
No. 12405119, the Natural Science Foundation of Sichuan Province under Grant No. 2025ZNSFSC0880, and Fundamental Research Funds for the Central Universities (Grant No. Y030242063002070). This work was also supported by the State Key Laboratory of Dark Matter Physics.

\newpage
\bibliography{paleoDM}

\end{document}